'Transcultural Robotic Nursing- An introduction to culturally-aware robots'

A recent analysis of global public health data found that people live longer with many more medical problems (Newton *et al* 2015). According to a King's fund report (2011), in the UK alone, the population of those between 65 to 84 years of age is projected to increase by 39% over the 30-year period from 2012 to 2032 and the number of older adults with health care needs is estimated to increase by more than 60%. In addition, epidemiologists predict that many older adults will be living on their own, despite their health care problems and of course they will require formal care. There are more than 11 million adults of 65 years old and over, that are currently living in the UK. It is estimated that more than two fifths of the NHS budget are being used for the care of older adults who live with complex medical problems (Mortimer and Green 2015).

The social care system in the UK is facing tremendous pressures especially in regard to the recruitment and retaining of carers. Stakeholders are using all possible options such as recruitment of health care professionals from overseas as well as adjusting the co-ordination and care delivery systems. In addition, considering the recent documented lapses of care in the NHS (Francis 2013) the assumption that every human carer is providing professional, high quality, compassionate care is under scrutiny.

Assistive robots (AR) have been defined as robots that provide support and help to a human user. Socially assistive robots (SAR) are sophisticated robots that provide assistance through social interaction, using speech, movements, gestures etc. (Feil-Seifer and Mataric´ 2005). Assistive robots can help fostering the independence and autonomy of older persons in many ways, by reducing the days spent in care institutions and prolonging the time spent living in their own home. Socially assistive robots are currently used in many settings and in healthcare but attitudes over the use of robots, especially in the care of elderly, vary.

According to a 2015 survey across Europe, 51% of respondents did not approve the use of robots in the care of older adults or as companions (Special Eurobarometer 2015). Similarly, the debate among healthcare professionals regarding the use of robots in the care of older adults is thriving. Recent experiments with the robot pet seal named Paro have shown to improve older adults' brain activity and lower their stress levels (Kachouie 2014; Leite 2013; Wada 2008) but the main fear and concern is that robots will replace and / or limit human interaction with older adults (Laitinen *et al* 2016; Walter 2017).

      Broadbent *et al* (2009) in a review of the literature related to the acceptance of health care robots among older adults, found many individual factors that significantly influenced their acceptance such as age, education, gender, experience with technology and culture. Older adults who perceived a need for an assistive device in order to maintain their independence were more likely to accept the technology but the fit between the robot's capabilities and the user's needs was very important (Frennert *et al* 2017). Another significant factor was culture with not all individuals feeling comfortable touching the robots and with Americans being, in general, more acceptant of the robots. When the robot could not understand the user's language the desire to interact and talk to the robot was limited (Broadbent *et al* 2009). A number of studies explored the differences in the acceptance of robots across different cultures, and found that people from different cultures not only have different preferences concerning how the robot should be and behave (Evers *et al* 2008) but also preferred robots who could better comply with the social norms of their own culture, in aspects such as the verbal (Andrist *et al* 2015; Rau *et al* 2009; Wang *et al* 2010;), and non-verbal behaviour (Trovato *et al* 2016) and the interpersonal distance (Eresha *et al* 2013; Joosse *et al* 2014). This preference does not merely affect the likeability of the robot. In a series of experiments on the influence of culture on Human Robot Interaction, participants from the US and China were asked to solve a task with the possibility of relying on the

suggestions of a robot assistant (Evers *et al* 2008). Experimenters analysed the level of trust, comfort, compliance, sense of control and anthropomorphism inspired by the robot on the people, and found that not only US and Chinese participants had different preferences, but each group had more trust and a more effective interaction with the robot complying with the norms of their culture (Wang *et al* 2010). Robots, along with sensors and telemedicine were identified as three technologies that can assist and prolong independent living among older adults with robots especially being used in the prevention of social isolation and depression. An experiment with Dutch participants and two robots, respectively customized for the German and Japanese culture, provides preliminary support to the hypothesis that acceptance of a robot could be directly proportional to cultural closeness (Trovato *et al* 2016). Despite the identification of 'culture' as an important factor in regards to the acceptability of robots among the elderly, 'culture' was not mentioned or addressed in a recent review exploring the effectiveness of different assistive technologies among the elderly (Khosravi and Ghapanchi 2016).

The importance of culture and the need for cultural competence in healthcare has been widely investigated in the nursing literature (Leininger 2002). Nurses acknowledge that one's culture has a significant impact on his/her health care decisions and for years they investigated the impact of cultural diversity on health and illness and they advocate that patient care should be culturally appropriate and competent. (Papadopoulos 2006).

We argue that if service robots are to be accepted in the real world by real people, they must thus account for the cultural identity and diversity of the assisted person and those of the healthcare team. Designers of personal robots for healthcare are faced with questions such as: "How should the robot greet a person?", "Should the robot ask direct questions or not?", "What distance should the robot keep from a person? Should it avoid or encourage physical contact?", "Is there any area of the house that it should consider off-limits?" Intuitively, the correct answer to all those questions is "It depends", and more precisely, it depends on the person's values, beliefs, customs and lifestyle, i.e., the person's culture, refined by personality and experiences (Sgorbissa *et al* 2017).

The authors are partners in the CARESSES (Culturally Aware Robots and Environmental Sensor Systems for Elderly Support) project which is a three year European - Japanese collaboration funded by the HORIZON 2020 programme and the Japanese Ministry of Internal Affairs and Communication. The project started in January 2017 with the ambition to develop culturally competent assistive robots. We define a culturally competent robot as a robot that knows general cultural characteristics, but it is aware that these general characteristics take different forms in different individuals, and is sensitive to cultural differences while perceiving, reasoning, and acting.

According to Sgorbissa *et al* (2017) today it is technically conceivable to build robots possibly operating within a smart environment populated with smart sensors and devices that reliably accomplish basic assistive services. However, state-of-the-art robots consider only the problem of "what to do" in order to provide a service: they produce rigid recipes, which are invariant with respect to the place, person and culture. It is now time that the technology addresses the question 'how to do' (Sgorbissa, *et al* 2017).

Well-grounded in the theoretical framework of cultural competence by Papadopoulos 2006, the CARESSES project is an innovative collaboration of nurses, computer scientists, psychologists, and artificial intelligence scientists aims to create a culturally competent and intelligent robot. The CARESSES team believes that cultural competence would allow assistive robots to increase user's acceptability by being more sensitive to their needs, customs and lifestyle, thus producing a greater impact on the quality of life of users and their caregivers, reducing caregiver burden, and improving the system's efficiency and effectiveness. CARESSES' culturally aware solutions are meant to expand the capabilities of any assistive or companion robot. They will be demonstrated on the Pepper robot, which is designed and marketed by Softbank Robotics, a partner of the project. The new culturally aware capabilities of the Pepper robot will include: a) communicating through culturally appropriate speech and gestures; b) moving independently; c) assisting the person in performing everyday tasks (e.g. reminding them to do a shopping list, suggesting menu plans, receiving visitors); d) providing health-related assistance (e.g. reminding the person to take her medication encouraging the person to engage in physical activity) ; e) providing easy access to technology (e.g. internet, video calls, smart appliances for home automation); f) providing entertainment (e.g. reading aloud, playing music and games etc ) and  g) promoting health safety through information and raising the alarm when needed.  During the last year of the project, the culturally aware robots will be tested in the UK and Japan among an elderly population. (www.caressessrobot.org )

In the CARESSES project, nurse researchers have the critical task in developing guidelines for Transcultural Robotic Nursing paving the way for future research in this field. Acceptability of robots in healthcare will continue to face challenges and the continuation of debate among health and social care professionals may lead to the appropriate use of smart technologies amongst different populations and different health care problems. The authors believe that is important for nurses to be involved and play a role in the development of future healthcare assistive robotic innovations and accept the challenge of increasing need and reduced human workforce in order to ensure that old and very old people in society continue to receive the care and support they need. At the same time, culturally competent assistive robots should be used ethically and must be considered as useful tools for human carers whom they will never replace but compliment (BSI 2016).

Word count (text only): 1,476 new word count 1,650

Transcultural Robotic Nursing    8References

Andrist S Ziadee M Boukaram H Mutlu B and M. Sakr M (2015) *Effects of culture on the credibility of robot speech: A comparison between English and Arabic.* HRI `15, Proceedings of the Tenth Annual ACM/IEEE International Conference on Human-Robot Interaction, 157–164.

BS 8611: (2016) *Robots and robotic devices. Guide to the ethical design and application of robots and robotic systems*. BSI Standards Publication.

Broadbent E Stafford R & MacDonald B (2009) Acceptance of healthcare robots for the older population: review and future directions. *International Journal of Social Robot*. 1, 319-330, doi 10.1007/s12369-009-0030-6.

Eresha G H¨aring M Endrass B Andr´e E & Obaid M (2013) *Investigating the influence of culture on proxemic behaviors for humanoid robots*. RO-MAN 2013, 22nd IEEE International Symposium, 430–435.

Evers V Maldonado H Brodecki T & Hinds P (2008) *Relational vs. group self-construal: untangling the role of national culture in HRI*. HRI `08, 2008, Proceedings of the 3rd ACM/IEEE International Conference on Human-Robot Interaction, 255–262.

Feil-Seifer D & Mataric´ M J (2005) *Defining Socially Assistive Robots*. Proceedings of the 2005 IEEE 9th International Conference on Rehabilitation Robotics, Chicago, IL, USA, 465-468.

Francis R (2013). *Report of the Mid Staffordshire NHS Foundation Trust Public Inquiry: executive summary*. London: Stationery Office, February 6th 2013. ISBN: 9780102981476. 116p. Presented to Parliament pursuant to Section 26 of the Inquiries Act 2005. Ordered by the House of Commons. HC 947.

Transcultural Robotic Nursing    9Frennert S Eftring H & Ostlund B (2017) Case report: implications of doing research on socially assistive robots in real homes. *International Journal of Social Robotics*, 1-15, doi 10.1007/s12369-017-0396-9.

Joosse M P Poppe R W Lohse M & Evers V (2014) *Cultural differences in how an engagement-seeking robot should approach a group of people*. CABS `14, Proceedings of the 5th ACM international conference on Collaboration across boundaries: culture, distance & technology, 121–130.

Kachouie R Sedighadeli S Khosla R & Chu M-T (2014) Socially Assistive Robots in Elderly Care: A Mixed-Method Systematic Literature Review. *International Journal of Human-Computer Interaction*. 30, 5, 369–393.

Khosravi P & Ghapanchi A H (2016) Investigating the effectiveness of technologies applied to assist seniors: a systematic literature review. *International Journal of Medical Informatics*. 85, 17-26.

Laitinen A Niemela M & Pirhonen J (2016) *Social Robotics, Elderly Care, and Human Dignity: A Recognition-theoretical Approach. What Social Robots Can and Should Do.* Seibt et al J (Eds.) IOS Press, 2016

Leininger M M (2002) *Transcultural nursing: concepts, theories, research and practice*. Third edition. McGraw-Hill, New York, NY.

Leite I Martinho C & Paiva A. (2013) Social Robots for Long-Term Interaction: A Survey. *International Journal of Social Robotics*. 5, 2 , 291–308.

Mortimer J & Green M (2015) *Briefing: the health and Care of older people in England 2015*. Age UK. http://www.cpa.org.uk/cpa/docs/AgeUK-Briefing-TheHealthandCareofOlderPeopleinEngland-2015.pdf

Newton et al (2015) Changes in health in England, with analysis by English regions and areas of deprivation, 1990–2013: a systematic analysis for the Global Burden of Disease